\documentclass[conference, 10pt]{IEEEtran}
\IEEEoverridecommandlockouts
\usepackage{cite}
\usepackage{amsmath,amssymb,amsfonts}
\usepackage{algorithmic}
\usepackage{graphicx}
\usepackage{textcomp}
\usepackage{xcolor}
\usepackage{mathrsfs}

\usepackage{caption, subcaption}
\def\BibTeX{{\rm B\kern-.05em{\sc i\kern-.025em b}\kern-.08em
    T\kern-.1667em\lower.7ex\hbox{E}\kern-.125emX}}
\begin{document}

\title{Predicting Survey Response with Quotation-based Modeling: A Case Study on Favorability towards the United States}

\author{
    \IEEEauthorblockN{Alireza Amirshahi, Nicolas Kirsch, Jonathan Reymond, Saleh Baghersalimi}
    \IEEEauthorblockA{École Polytechnique Fédérale de Lausanne (EPFL)
    \\\{alireza.amirshahi, nicolas.kirsch, jonathan.reymond, saleh.baghersalimi\}@epfl.ch}
}

\IEEEoverridecommandlockouts
\IEEEpubid{\makebox[\columnwidth]{----/23/\$31.00~\copyright2023 IEEE \hfill} \hspace{\columnsep}\makebox[\columnwidth]{ }}

\maketitle

\IEEEpubidadjcol

\begin{abstract}
The acquisition of survey responses is a crucial component in conducting research aimed at comprehending public opinion. However, survey data collection can be arduous, time-consuming, and expensive, with no assurance of an adequate response rate. In this paper, we propose a pioneering approach for predicting survey responses by examining quotations using machine learning. Our investigation focuses on evaluating the degree of favorability towards the United States, a topic of interest to many organizations and governments. We leverage a vast corpus of quotations from individuals across different nationalities and time periods to extract their level of favorability. We employ a combination of natural language processing techniques and machine learning algorithms to construct a predictive model for survey responses. We investigate two scenarios: first, when no surveys have been conducted in a country, and second when surveys have been conducted but in specific years and do not cover all the years. Our experimental results demonstrate that our proposed approach can predict survey responses with high accuracy. Furthermore, we provide an exhaustive analysis of the crucial features that contributed to the model's performance. This study has the potential to impact survey research in the field of data science by substantially decreasing the cost and time required to conduct surveys while simultaneously providing accurate predictions of public opinion.
\end{abstract}

\begin{IEEEkeywords}
Sentiment Analysis, Survey Response Prediction, Quotation-based Modeling, K-Nearest Neighbors
\end{IEEEkeywords}

\section{Introduction}
Surveys play a vital role in enabling organizations and governments to obtain data and insights from their target populations~\cite{groves2011survey}. They offer valuable information on consumer preferences, needs, and behaviors, which are essential for companies to create products and services that meet the customers' requirements. Surveys are also an indispensable means for governments to gauge public opinion and acquire data on significant national issues, such as healthcare, education, and employment. Moreover, surveys facilitate policymakers in evaluating the effectiveness of government policies and programs and identifying areas that require improvement. They serve as a foundation for evidence-based decision-making, enabling organizations and governments to make informed decisions based on reliable data. Therefore, surveys are a crucial tool for data science, providing the primary data required for analysis and modeling and assisting organizations and governments in obtaining insights that can guide strategic decision-making.

While surveys have many benefits, they also come with several challenges~\cite{dillman2014internet, yeager2011comparing, couper2002new}. One of the most significant problems with surveys is the cost associated with designing, conducting, and analyzing them. Surveys often require a considerable amount of time and resources to develop and administer, and analyzing the data can be time-consuming and costly. Additionally, surveys can suffer from incompleteness, meaning that not all participants may answer all questions, leading to missing data. Low response rates can also be a problem, which may affect the representativeness of the sample and lead to biased results. Moreover, during the COVID-19 pandemic, conducting surveys became more challenging, with restrictions on face-to-face interactions and a shift towards online surveys leading to potential issues with sample and data quality. 

The Pew Research Center~\cite{PEW}, based in the United States, conducts surveys to gather public opinion on various topics that impact the world. These encompass political matters, immigration, ethnicity, religion, the economy, and science. A focus of their research is the image and reputation of the United States in the eyes of the rest of the world. In their survey, individuals from around the world are asked to share their thoughts on American society and politics. The PEW Research Center releases the results of these surveys as articles after data analysis~\cite{wike2017us}.

\begin{figure}[t]
    \centering
    \includegraphics[width=\linewidth]{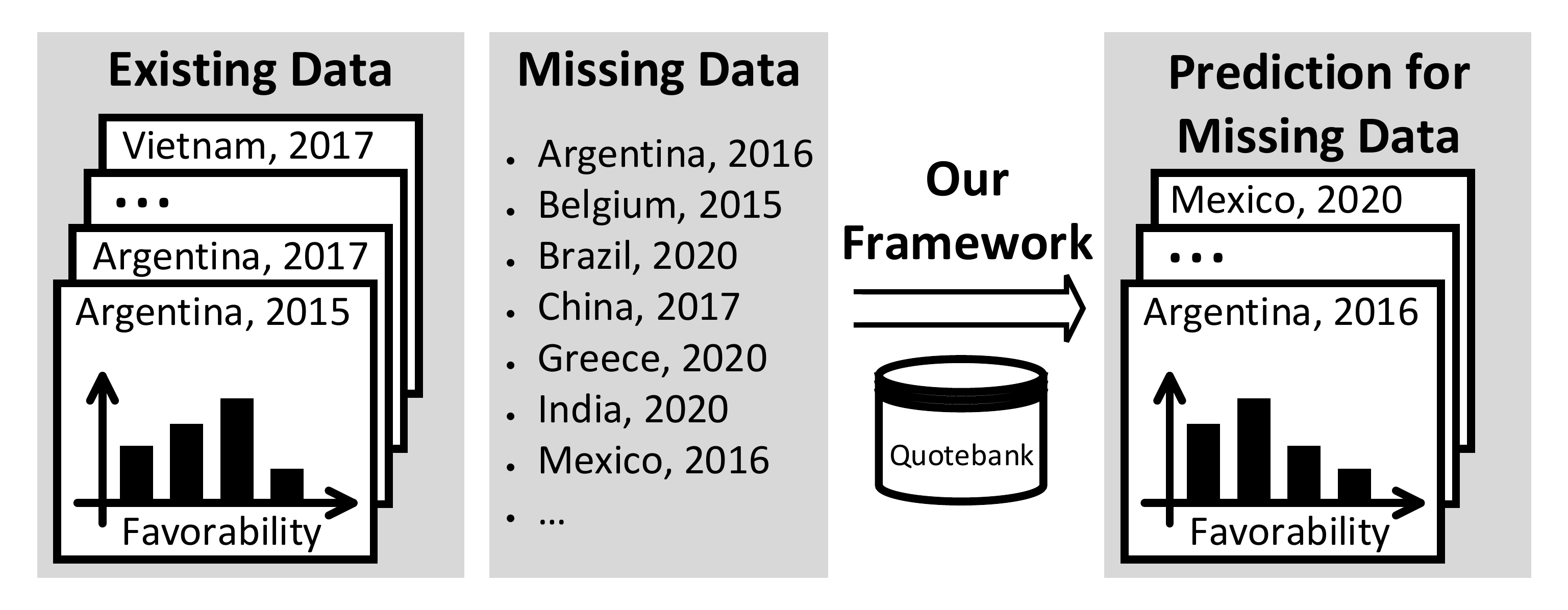}
    \captionsetup{font=small}
    \caption{Motivation for our approach to predict missing data. This figure outlines the temporal and geographical distribution of existing survey data in PEW dataset and identifies gaps where survey data is currently lacking (missing data) across various countries and years.}
    \label{fig:motivation}
\end{figure}

In this paper, we propose a solution to the challenges faced in survey research by predicting survey responses for countries and years where data is incomplete or unavailable, without the need for conducting an actual survey. Our motivation, illustrated in Figure~\ref{fig:motivation}, originates from the observation of \emph{existing data} in PEW dataset, representing specific countries and years for which surveys have been conducted. Conversely, there are \emph{missing data} for several countries or years where surveys have not been carried out. Our goal is to predict the actual sentiment of these missing populations. This is achieved by performing sentiment analysis on the Quotebank dataset\cite{vaucher2021quotebank}, a quotation database comprised of quotes from various news sources published between 2008 and 2020.

We first gather essential details about the speakers, including their name, origin, and the time the quotes were made from the QuoteBank. We identify keywords to extract quotes relevant to the United States and perform sentiment analysis on each quote. To enhance the precision of our results, we train a K-Nearest Neighbor~(KNN) model to convert the continuous distribution generated by our sentiment analysis into the discrete distribution found in the PEW dataset. Therefore, we are able to predict the results of the PEW analysis regarding the favorability of the United States in each country and across various points in time.

We present a novel method for predicting survey responses by utilizing quotations through machine learning techniques. Our study focuses on evaluating the degree of favorability towards the United States and contributes to the field of data science in several ways:
\begin{itemize}
    \item This paper is the first, to the best of our knowledge, to use multiple data sources, such as PEW and Quotation results, for the purpose of predicting survey response.
    \item Our approach enables us to predict survey responses for countries that have never been studied in the survey, as well as for countries that have been studied but not in specific years.
    \item We propose a KNN regressor to predict the discrete distribution of responses based on a continuous distribution of quotation sentiment.
    \item This work proposes a keyword extraction and media bias reduction methodology on Quotebank to improve the reliability of our results and generate a dataset that is more comparable to real data.
\end{itemize}

The organization of the rest of this article is as follows: In Section~\ref{sec:RelatedWork}, we outline the problem of interest and examine previous research. Section~\ref{sec:Methodology} details the training of our proposed approach for estimating the United States favorability. Section~\ref{sec:ExperimentalSetup} covers the dataset and experimental design. In Section~\ref{sec:Result}, we assess our results and evaluate the advantages of our proposed method. Finally, Section~\ref{sec:Conclusion} summarizes the essential findings and conclusions of the study.

\section{Related Work}
\label{sec:RelatedWork}

Several studies have explored the use of language processing algorithms to predict survey responses based on the semantic properties of survey items. For instance, a study~\cite{arnulf2014predicting} investigated the predictability of survey responses based on semantic properties of survey items, finding that language processing algorithms could explain a significant portion of the variance in response patterns.  Another study~\cite{axhausen2010predicting}  aimed to develop an efficient and objective method for rating response burden in mail and mail-back surveys, which would help market research firms estimate response rates in advance. Moreover, a study~\cite{han2019using} identified factors that predicted response rates in consumer satisfaction, loyalty, and trust surveys, discovering associations with data collection mode, survey sponsors, privacy,  and cultural orientation. Lastly, a more recent study~\cite{hamdan2022predicting} used machine learning to predict customer loyalty in the hotel industry, finding that customer satisfaction, brand image, and service quality were the most important factors. While the aforementioned works have made strides in the field of survey studies, their approaches largely revolve around analyzing semantic properties of survey items, or predicting survey response rates. Conversely, our work  introduces a more generalized approach that is not limited to the boundaries of traditional surveys, allowing us to predict responses even in situations where conventional survey data is incomplete or unavailable.

Quotations have become a valuable source of information in the digital era. Several studies have been conducted on the application of quotations in data science, including the development of corpora and datasets, recommendation models, and extraction and attribution methods. A noteworthy contribution to the study of quotations in data science is Quotebank~\cite{vaucher2021quotebank}, which provides an extensive open corpus of 178 million quotations extracted from 162 million English news articles published between 2008 and 2020. In addition to the corpus, the authors introduce Quobert, a framework for extracting and attributing quotations from large corpora with minimal supervision. Quobert utilizes a bootstrapping approach that avoids the need for manual labeling by deriving training data from a single seed pattern and leveraging the corpus redundancy to fine-tune a BERT-based model~\cite{devlin2018bert}.

In the context of quote recommendation, the QuoteR~\cite{qi2022quoter} dataset was built to facilitate research on the task of recommending quotes.The dataset, which is larger than previous unpublished counterparts, consists of three parts: English, standard Chinese, and classical Chinese. To assess the effectiveness of existing quote recommendation methods, an extensive evaluation was conducted on QuoteR, leading to the proposal of a new model that significantly outperforms previous methods across all three parts of the dataset. These findings highlight the potential of quote recommendation methods in enhancing writing quality by efficiently identifying appropriate quotations that can make written work more persuasive and polished.

An important application of quotations is attribution recognition, and the Political News Attribution Relations Corpus 2016 (PolNeAR)~\cite{newell2018attribution} is the one of the most comprehensive attribution relations corpus. This dataset includes high-quality attribution annotations and overcomes the limitations of existing resources by incorporating more diverse publishers. PolNeAR was constructed from news articles covering the political candidates during the year leading up to the United States Presidential Election in November 2016. The dataset has the potential to support the development of advanced end-to-end solutions for attribution extraction and to encourage interdisciplinary collaboration between natural language processing~(NLP), communications, political science, and journalism communities. In addition to the dataset, the authors contributed updated guidelines aimed at enhancing clarity and consistency in the annotation task, as well as an annotation interface customized to the task.

In conclusion, quotations have been applied in various data science tasks. These applications demonstrate the usefulness of quotations in extracting valuable information from textual data, improving the quality of scientific news articles~\cite{smeros2019scilens}, and analyzing character relationships in novels~\cite{jia2020speaker}.

\section{Methodology}
\label{sec:Methodology}

\begin{figure*}[t]
    \centering
    \includegraphics[width=0.96\linewidth]{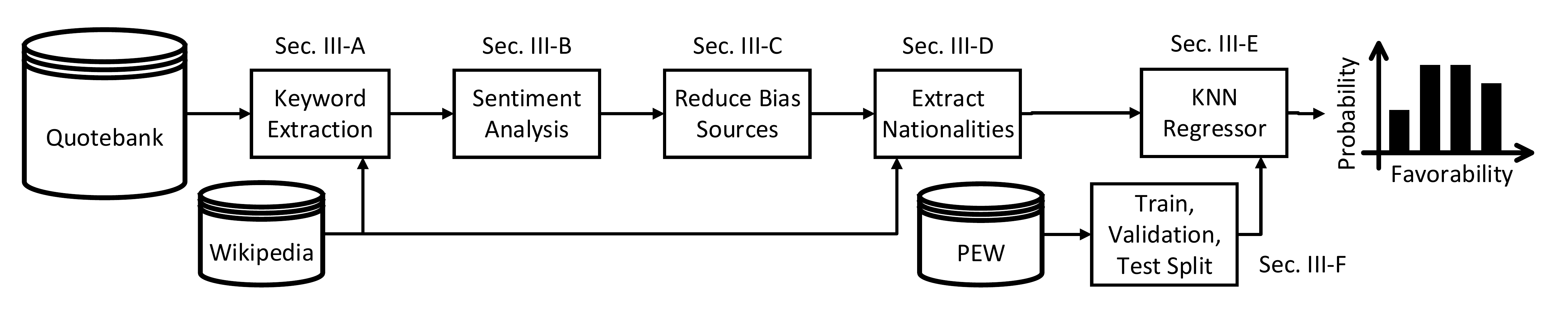}
    \captionsetup{font=small}
    \caption{Overview of the proposed quotation-based predictive modeling method for assessing favorability towards the United States. This method offers a novel approach to predicting survey responses by leveraging multiple data sources and advanced techniques.}
    \label{fig:overview}
\end{figure*}

In this paper, we propose a quotation-based predictive modeling method to assess favorability towards the United States. The structure of our proposed method is shown in Figure~\ref{fig:overview}, which consists of five main steps. First, we collect data from the Quotebank database and extract all the keywords related to the survey question, which in our case is about United States favorability. Second, we use a sentiment analysis model to analyze and assign a degree of positive or negative sentiment to each quote. Third, we perform a bias reduction step to eliminate quotes from biased media outlets. Fourth, we extract the nationality of the speakers using the Wikipedia database. Finally, we use a KNN regressor to predict the survey response distribution based on the quotes and sentiment analysis results to make it similar to the PEW database.

In the following sections, we describe each step of our proposed method in detail. We explain the data sources and algorithms implemented, including the specific sentiment analysis model and the KNN regressor. We also provide a thorough evaluation of our method and its performance for predicting survey responses. Our proposed method offers a novel approach to predicting survey responses by leveraging multiple data sources and advanced techniques. We believe it has the potential to improve the accuracy and reliability of public opinion research significantly.

\subsection{Keyword extraction}

Quotebank is an open corpus providing an extensive resource for various NLP tasks such as quotation attribution, sentiment analysis, and topic modeling. In this work, we considered the Quotebank database from 2015 to 2020, which contains 18.7 GB of data. At first, we attempted a manual approach to extract data containing keywords relevant to our objective. We defined the keywords "US," "U.S.," "USA," and "United States" and extracted 253.1 MB of data containing these keywords. However, the number of quotes obtained was not adequate for our analysis. Therefore, we enriched our keyword list using a different approach. We extracted all American speakers from the Wikipedia database and determined which ones were present in the Quotebank dataset. We discovered the most frequent American speakers and created enriched keywords such as their last or complete names. By filtering all the quotes containing one of the total keyword lists, we obtained a more significant number of relevant quotes for our analysis.

Overall, our keyword extraction methodology involves a combination of manual and automated approaches, where we extract relevant data from the massive Quotebank corpus and enrich our keyword list using external sources.

\subsection{Sentiment Analysis}
Sentiment analysis is a widely used technique for determining the polarity of text, ranging from positive to negative. In our study, we applied sentiment analysis to the quotes extracted from Quotebank to determine the overall sentiment towards the United States. The sentiment analysis task involves extracting features from the raw text, such as n-grams or other contextual clues, and using machine learning algorithms to classify the sentiment.

We used a pre-trained sentiment analysis model to analyze the sentiment of the quotes extracted from Quotebank. The model takes raw text as input and outputs a probability distribution over positive and negative sentiment. We applied the model to each quote individually and recorded the predicted sentiment degree ranging from $-1$ to $1$.

\subsection{Reducing bias: Media sources}

Reducing bias in our method is crucial to ensure the consistency and reliability of our analysis. If our data is biased, our results will be skewed, leading to incorrect conclusions. Therefore, taking steps to identify and reduce bias in our methodology is essential for producing high-quality and trustworthy research.

One of the primary sources of bias in the media is the selection of quotes to fit a particular narrative or point of view. While attempting to be impartial, articles written by humans can still carry some inherent bias. This bias could potentially result in including quotes that align with a specific viewpoint while discarding opposing quotes, leading to a biased quotation corpus. Media outlets with clear political, social, or geopolitical views may also have their own selection processes for quotations, leading to a biased dataset. Some media outlets may selectively include only positive or negative quotes about the United States, leading to extreme sentiment scores.

To mitigate this bias, it is important to identify media outlets that exhibit such behavior and eliminate them from the dataset. This step can lead to better data for analysis, transforming the analysis from "how favorable are countries to the United States as seen by the media?" to "how favorable are countries to the United States?" 

To determine the extent of bias in media sources, we extracted the sentiment distribution of the 30 most common sources. Any media outlet that was significantly different from others (p-value $<$ 0.05) was eliminated based on differences in the distribution of sentiment. By removing such sources, we aim to increase the validity of our results and generate a dataset that is more comparable to real data.

\subsection{Extract Nationalities}
In order to investigate the favorability towards the United States based on nationalities, we need to know the nationalities of the speakers for each quote in the Quotebank. To extract this information, we utilized a Wikipedia database. However, it should be underscored that not all quotes in the Quotebank have identifiable speakers. Since the primary goal of this project is to ascertain the sentiment of a specific country towards the United States, it is essential to establish the speaker's nationality. Thus, quotes where the speaker's nationality is undetermined are considered unfit for our analysis and are consequently removed. Additionally, some speakers may have dual nationalities, so we have counted the quotes for both countries. With this information, we can further explore how sentiment towards the United States may vary among different nationalities.

\subsection{KNN Regressor}

KNN is a popular machine learning algorithm for both classification and regression tasks. It is a non-parametric method that stores all available cases and uses them to make predictions or decisions based on similarity measures. The KNN algorithm assumes that similar things exist in close proximity, and therefore, it predicts a label or value for a new data point based on the closest labeled data points in the training set. The KNN algorithm is simple, easy to implement, and often provides good results with small or medium-sized datasets.

\begin{figure}[t]
    \centering
    \includegraphics[width=\linewidth]{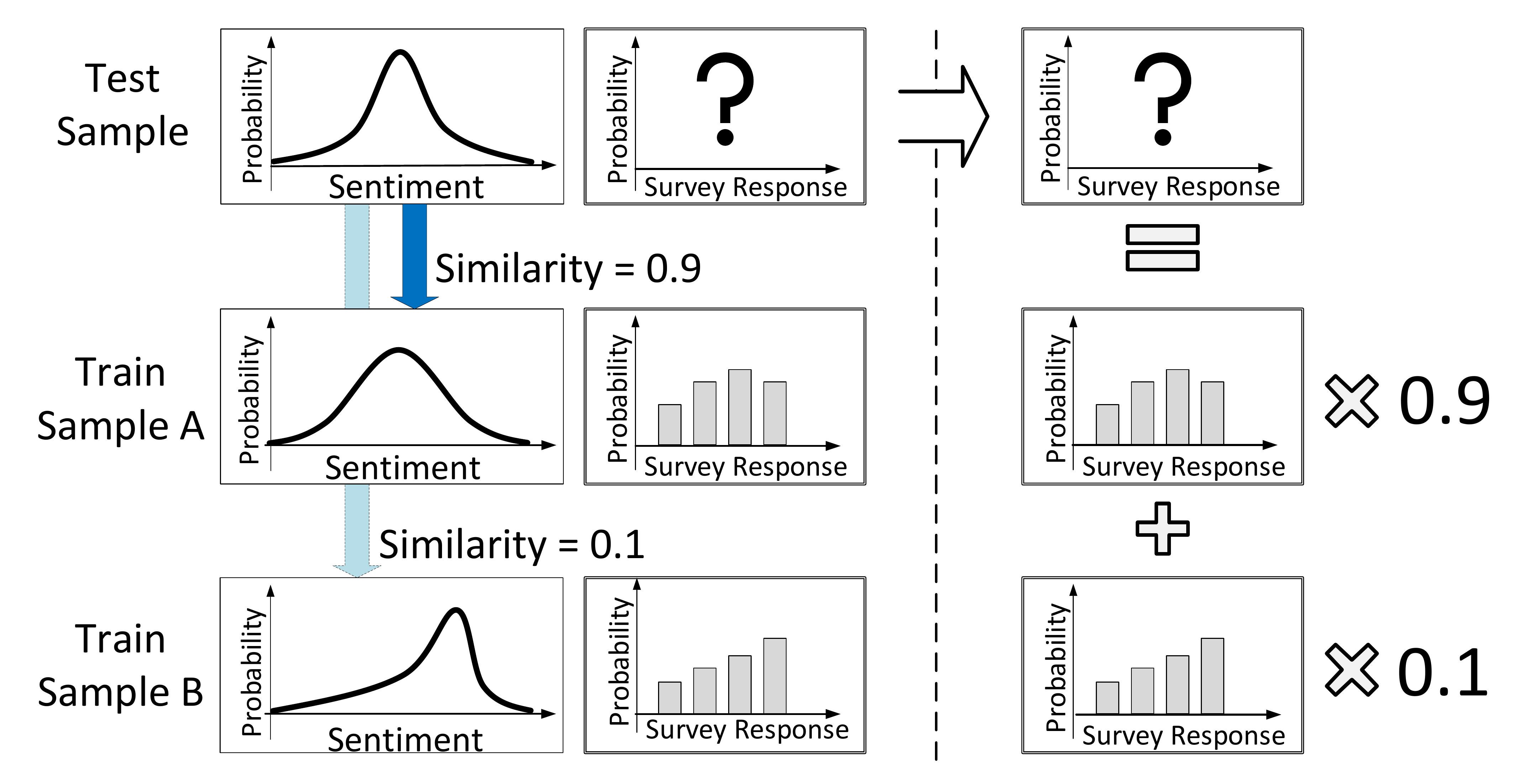}
    \captionsetup{font=small}
    \caption{Illustration of KNN Regression for Quotation-Based Modeling for predicting survey responses. The sentiment distributions of two training samples and one test sample, obtained from Quotebank, are shown as probability density functions. The goal is to predict the survey response for the test sample, which is done by computing the similarity of its sentiment distribution to the sentiment distributions of the K nearest neighbors in the training samples and performing a weighted average of their survey responses.}
    \label{fig:knn_reg}
\end{figure}

Our paper utilizes the KNN regression technique, which leverages KNN to forecast the output value for a continuous variable. The sentiment data for each sample is extracted from Quotebank, as shown in Figure~\ref{fig:knn_reg} in the form of a probability density function. While the test sample lacks an output, the train samples have survey responses in the form of a discrete probability density function extracted from PEW. The goal is to predict the survey response for the test sample using the train samples.

To accomplish this, we first compute the similarity of the sentiment distribution for the test sample to that of the train samples. Specifically, we measure similarity as the normalized inverse of KL divergence between the sentiment distribution of the test sample and that of the train samples. KL divergence is a measure of the difference between two probability distributions, and it is calculated as:

\begin{equation}
D_{KL}(P||Q)=\sum_{i}P(i)\log\frac{P(i)}{Q(i)},
\end{equation}
where $P$ and $Q$ are two probability distributions. As shown in Figure~\ref{fig:knn_reg}, a lower KL divergence value indicates greater similarity between the distributions. We assume that the higher the similarity, the more likely the survey response of the test sample similar to the training sample. We leverage this similarity score to predict the survey response for the test sample. Specifically, we perform a weighted average of the survey responses for the K nearest neighbors of the test sample in the training samples, where K is a predetermined value. For instance, in Figure~\ref{fig:knn_reg}, we define K as 2. We assign weights to each survey response based on the similarity scores between the sentiment distributions of the test and train samples. By this method, we predict the survey response for the test sample.

In choosing the KNN regression technique for our study, we considered the size and complexity of our dataset and the nature of our task. Despite its richness in detail, our dataset is limited in volume and diverse across various nations, raising concerns about overfitting with optimization-based regressors. KNN offers a robust alternative, excelling in leveraging local information, resisting overfitting, and providing interpretable results – key assets for our sentiment analysis.

In summary, our proposed KNN regression method involves computing the similarity between the sentiment distributions of the test and train samples, and using this similarity score to predict the survey response for the test sample. We perform a weighted average of survey responses for the K nearest neighbors of the test sample, and assign weights based on the similarity scores. This method allows us to make accurate predictions of survey responses, as demonstrated in Figure~\ref{fig:knn_reg}.

\subsection{Real-World Scenarios for Survey Response Prediction}

In this section, we investigate two real-world scenarios based on the availability of data in the reference dataset, which in our case is PEW. We split the data into the train set, validation set, and test set, depending on the scenario. The two scenarios are described as follows:

\begin{figure}[ht]
    \centering
    \begin{subfigure}{0.23\textwidth}
    \centering
    \includegraphics[width=\linewidth]{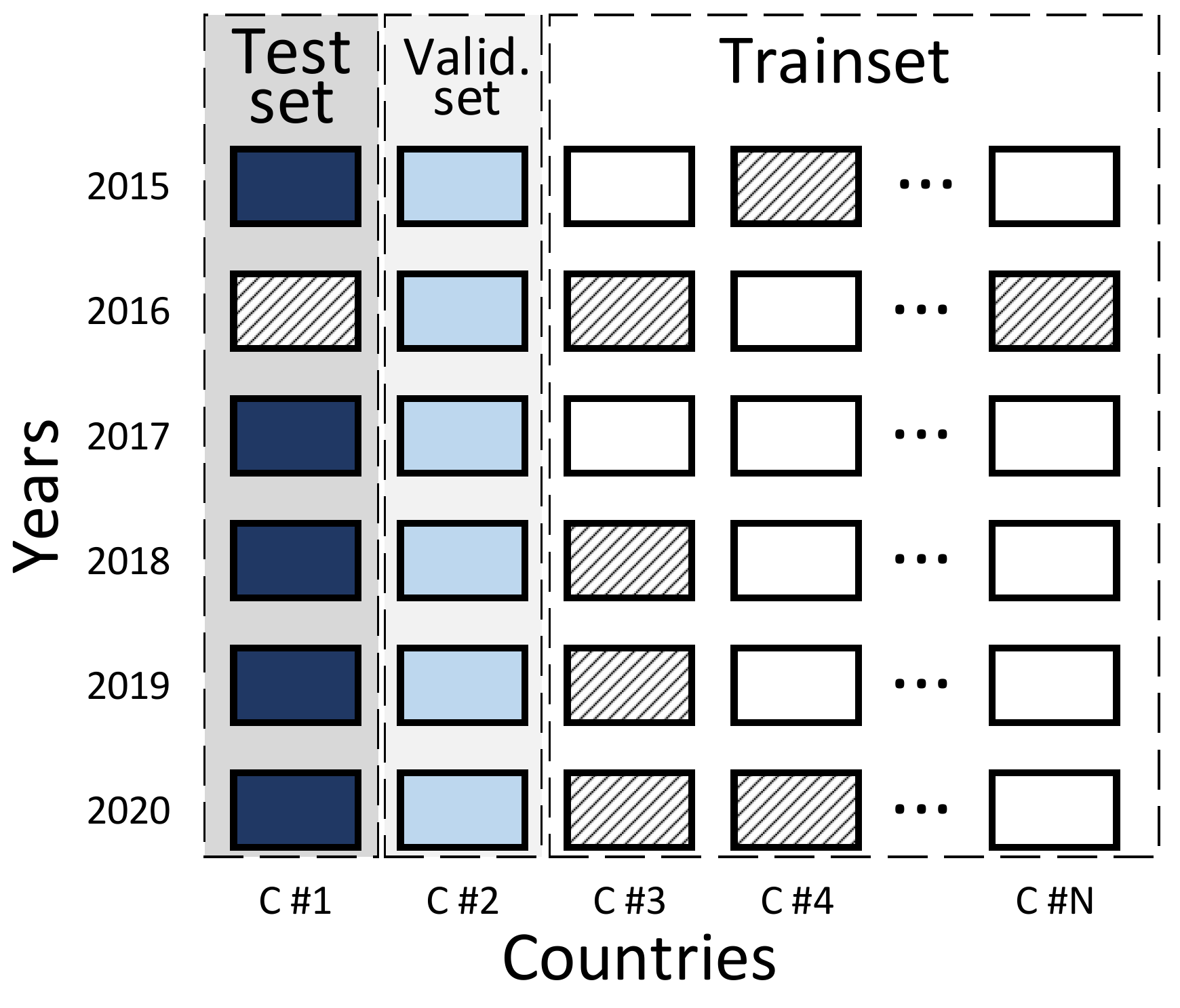}
    \caption{}
    \label{fig:loco_approach}
\end{subfigure}
\begin{subfigure}{0.225\textwidth}
    \centering
    \includegraphics[width=\linewidth]{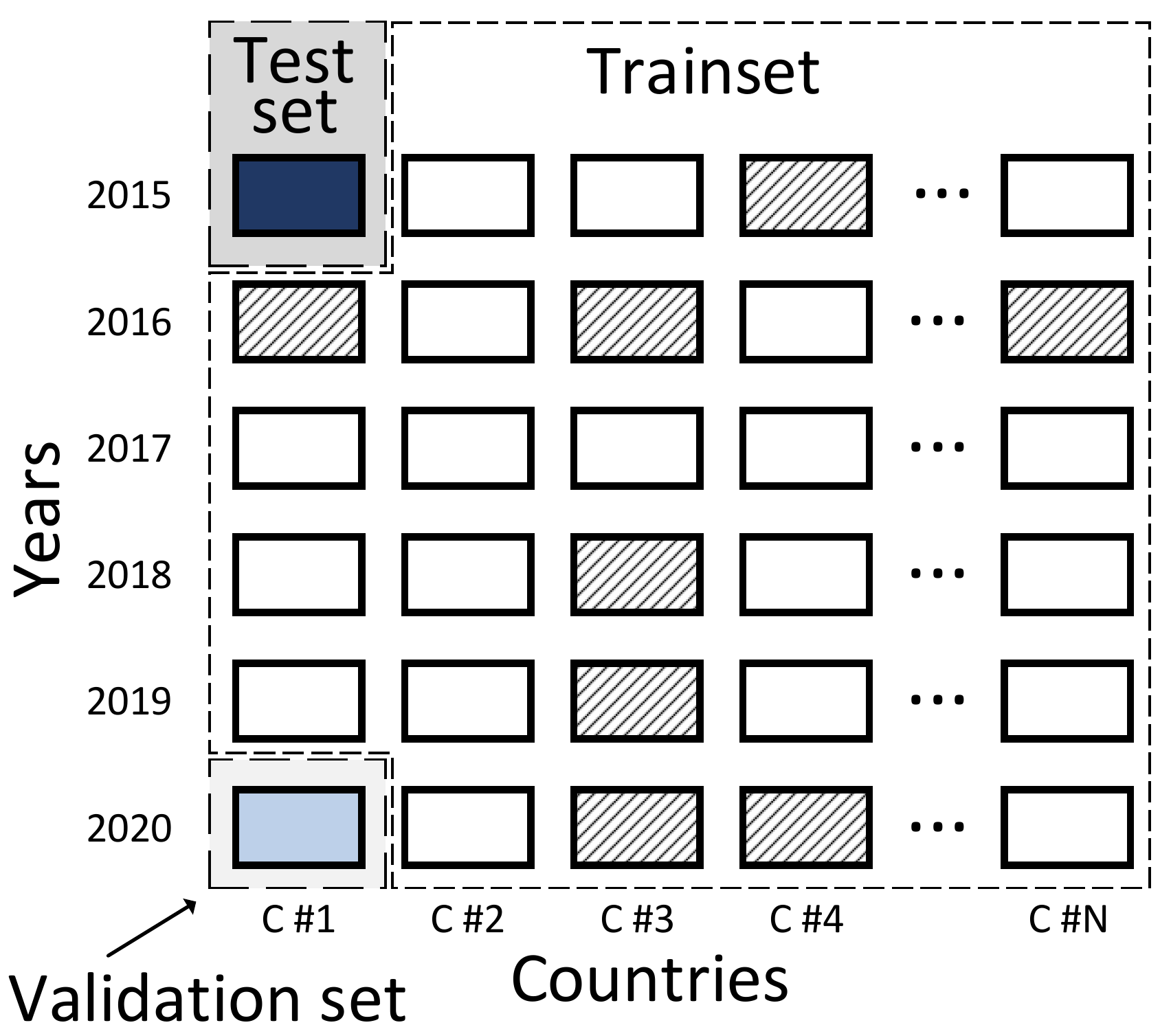}
    \caption{}
    \label{fig:scv_approach}
\end{subfigure}
\captionsetup{font=small}
\caption{Comparison of Two Quotation-Based Modeling Scenarios. Figure (a) illustrates the leave-one-country-out~(LOCO) scenario, while Figure (b) illustrates the same-country-validation~(SCV) scenario. In both figures, the test set is indicated by a darker color, the validation set by a lighter color, and the training set by white. The diagonal stripe patterns show the years that are missed from the original PEW dataset.}
\label{fig:scenarios}
\end{figure}

\subsubsection{Leave-One-Country-Out Scenario~(LOCO)} 
This scenario assumes that there is no data in the PEW dataset for the target country. This can occur due to various reasons such as limited resources, restricted access, or lack of interest in that country. Figure~\ref{fig:loco_approach} depicts this scenario and the corresponding data split. We select one country as the test set and use the rest of the countries as the training set. For the validation set, we choose one country from the remaining countries and use all the available years for that country. We employ KNN algorithm to find the nearest neighbors of each sample in the validation set from the training set. We compute the loss based on the KL divergence of the predicted survey response for validation with the PEW survey response in the training set for K values ranging from 1 to 9. The loss function for a country $c_i$ in the validation set, with the number of nearest neighbors as $k$ is calculated as \eqref{eq:loss_val}.

\begin{equation}
    \mathcal{L}^{k}_{c_i \in \text{val}} = \sum_{y \in Y_i} \text{KL} (\text{PEW} (c_i, y) || \text{Predict} (c_i, y) ; K=k)
    \label{eq:loss_val}
\end{equation}
Where $Y_i$ denotes the years included in the PEW dataset for the country $c_i$. We repeat this process by rotating the validation set to other countries and calculate the loss. This approach is N-fold cross-validation. We find the total loss corresponding to $k$ number of nearest neighbor as \eqref{eq:loss_total}.

\begin{equation}
    \mathcal{L}^{k} = \sum_{c_i \neq c^*} \mathcal{L}^{k}_{c_i}
    \label{eq:loss_total}
\end{equation}
Where $c^*$ represents the target country which is the test set and we are not using any data from this country during the training. We select the best K as the hyper-parameter and use this specific K to predict the output of the test set as \eqref{eq:best_k}.
\begin{equation}
    \text{Best}\;k = \operatorname*{argmin}_k \mathcal{L}^{k}
    \label{eq:best_k}
\end{equation}

 Finally, we repeat the whole process for all the countries in the dataset to simulate every single country as a test case.

\subsubsection{Same-Country-Validation Scenario~(SCV)} In this Scenario, we aimed to evaluate the effectiveness of our proposed method in cases where the target country has some data available in PEW but not for all the years of interest. This scenario can be seen as a more challenging case compared to the LOCO scenario. Regarding the data split, as shown in Figure~\ref{fig:scv_approach}, we select one specific year from the target country as the test set and another year from the same country as the validation set. For the training set, we use the samples from all other countries and the remaining years of the target country. Similar to the LOCO scenario, we apply the KNN algorithm and calculated the loss for different values of K from 1 to 9 for the validation set. We repeat this process by rotating the validation set to other years of the same target country. In this scenario, we multiply a constant $\lambda > 1$ to the similarity between two samples from the same country to increase the impact of results from the same country from a different year. We obtain this constant using cross-validation. 

The use of these scenarios helps us to analyze and evaluate the performance of our model in different situations. By considering the data availability and selecting appropriate scenarios, we can improve the effectiveness of our model and increase its accuracy. 


\section{Experimental Setup}
\label{sec:ExperimentalSetup}

\subsection{PEW Dataset}
\begin{figure*}[t]
    \centering
    \includegraphics[width=0.96\linewidth]{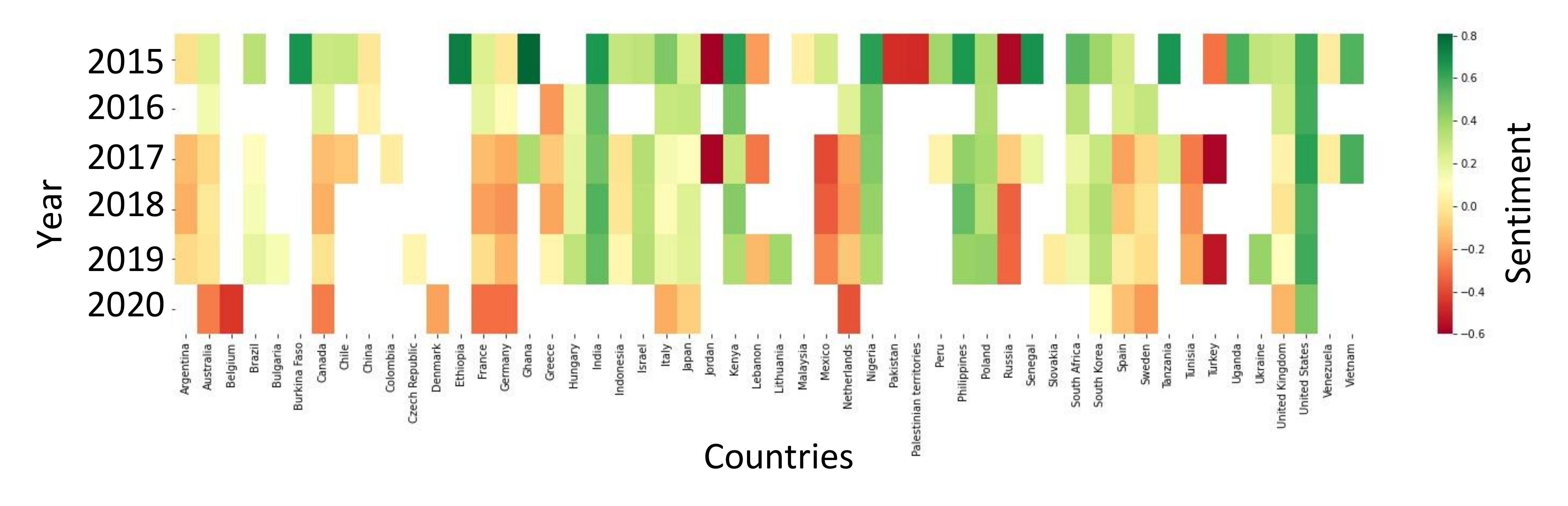}
    \captionsetup{font=small}
    \caption{Average survey response for the question ``Please tell me if you have a very favorable, somewhat favorable, somewhat unfavorable, or very unfavorable opinion of the United States'' across 52 countries from 2015 to 2020. Values of 1, 0.5, -0.5, and -1 were assigned to very favorable, somewhat favorable, somewhat unfavorable, and very unfavorable, respectively. White cells indicate years where no data was collected.}
    \label{fig:pew_heatmap}
\end{figure*}

This study uses the PEW database, which was collected from surveys conducted in multiple countries. The database contains various questions answered by the participants, and in this work, we focus on one specific question: ``Please tell me if you have a very favorable, somewhat favorable, somewhat unfavorable, or very unfavorable opinion of the United States''.  The average survey response to this question for each country and year is shown in Figure~\ref{fig:pew_heatmap}. We represent the responses to this question as numerical values of 1, 0.5, -0.5, and -1 for very favorable, somewhat favorable, somewhat unfavorable, and very unfavorable, respectively. The data covers 52 countries and 172 cells of (country, year) in total, with an average of 1007 participants $\pm$ 379 in each (country, year).

To ensure that the results of this study reflect the global image of the United States from outside, we excluded the United States itself from the analysis. Additionally, we removed three other cells from the dataset because there were not enough quotations in the Quotebank dataset for the corresponding year and country. This reference database serves as the basis for our proposed quotation-based modeling approach to predict survey response towards the United States.

\subsection{Sentiment analysis tool}
Pre-trained language models have proven to be effective for sentiment analysis, but they are usually trained on general text corpus, which may not include quotations. Retraining the models on quotation-specific data requires a significant amount of labeled data, which may not be feasible in many cases due to the high cost and time involved. Therefore, there is a need for a sentiment analysis tool that can accurately process quotations without the need for retraining the models. 

 This work aims to address this issue by evaluating the performance of pre-trained sentiment analysis models on quotes. To achieve this, a random sub-sample of quotes was extracted from Quotebank and assigned a sentiment score (-1, -0.5, 0, 0.5, 1) by the authors of this work. Three different tools, namely NLTK~\cite{hardeniya2016natural}, Flair~\cite{akbik2019flair}, and TextBlob~\cite{loria2018textblob}, were then tested on the sentiment scores and their performance was compared. 

While there was not a significant difference in performance among the sentiment analysis tools tested on the Quotebank sub-sample, Flair stands out as a potential choice, especially considering the high variance caused by the small test set size. It is worth noting that Flair utilizes neural networks for processing, which rules out the use of GPUs for scaling up to larger datasets. Flair is a leading NLP library that offers pre-trained models for various NLP tasks, including sentiment analysis, named entity recognition, and part-of-speech tagging. Its flexibility, ease of use, and high performance make it a popular option for both academic and industry use cases. The sentiment analysis model in Flair is highly accurate and effective, having achieved state-of-the-art results on several benchmark datasets. By utilizing Flair in our analysis of Quotebank, we gained valuable insights into the sentiment expressed towards the United States in the collection of quotations.

\subsection{Biased media}
\label{sec:bias}

\begin{figure}[ht]
    \centering
    \includegraphics[width=0.8\linewidth]{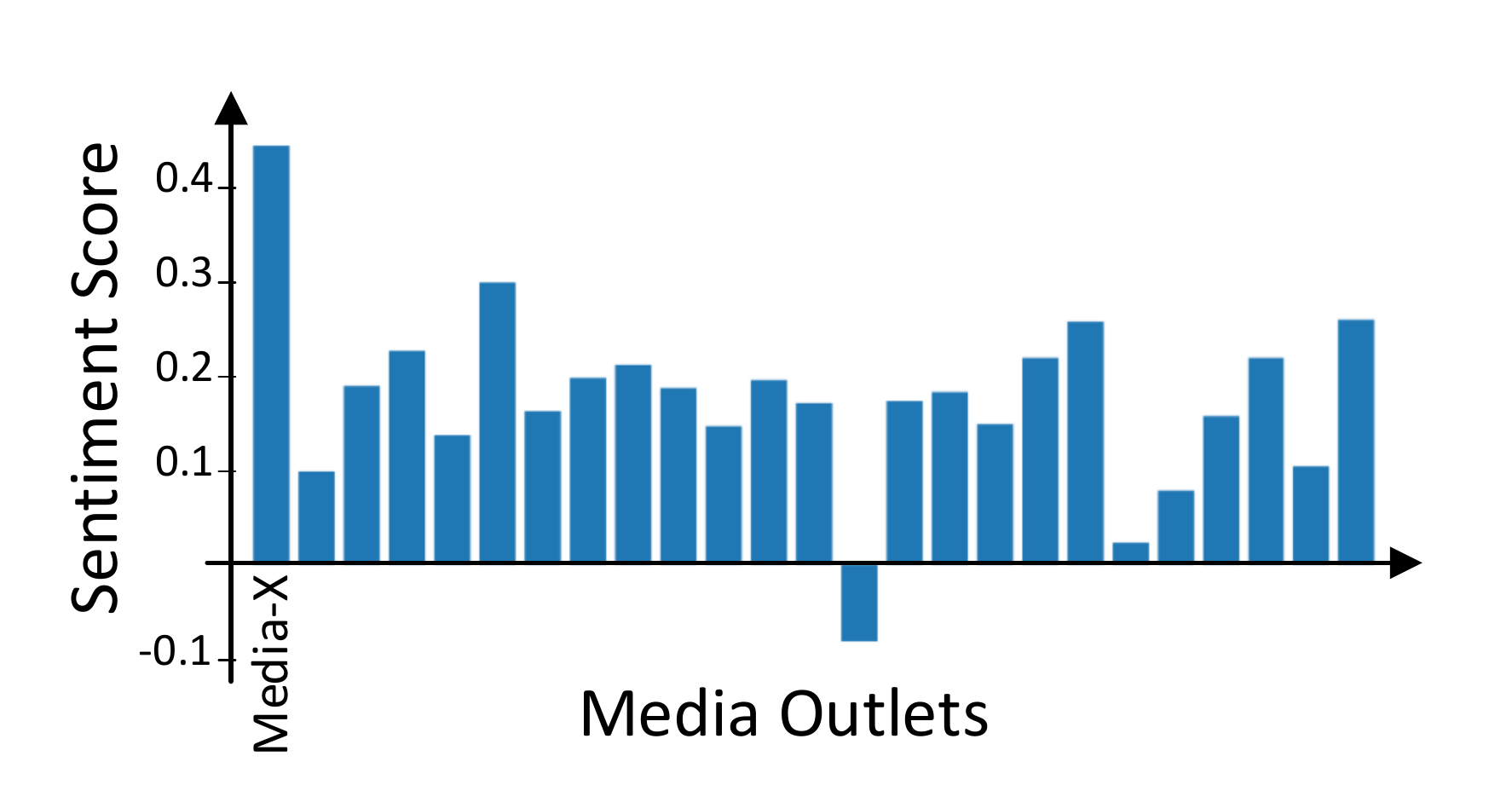}
    \captionsetup{font=small}
    \caption{Barplot comparing the sentiment scores of different media outlets based on their quotation sentiment towards the United States. To preserve the privacy, we have removed the name of the media outlets in this figure.}
    \label{fig:mean_sent}
\end{figure}

In our analysis of media sentiment scores, as shown in Figure~\ref{fig:mean_sent}, we observed that most media outlets had sentiment scores in the vicinity of 0, suggesting a neutral attitude towards the United States. However, one media outlet, stood out with a much higher sentiment score beyond 0.4, indicating a different perception of sentiment towards the United States. For privacy reasons, the name of this media outlet is not disclosed in this paper, and we refer to this media outlet as Media-X in this section.

To further explore this distinction, we conducted a statistical analysis using a t-test between the distribution of Media-X sentiment scores and the distribution of sentiment scores of all other media outlets. The resulting p-value was $< 0.05$, indicating a statistically significant difference between this specific media and the other media outlets.

Our findings suggest that Media-X does not represent the same distribution of quotation sentiment as the other media outlets. Its higher mean sentiment score indicates a positive bias towards the United States. Therefore, filtering out Media-X from the dataset could lead to improved reliability and a dataset more comparable to the PEW data.

The observation of Media-X distinct sentiment score distribution highlights the importance of analyzing and filtering biased media outlets in sentiment analysis. Our approach of using a bias reduction step in the quotation-based predictive modeling method proposed in this paper could help improve the reliability of public opinion research by reducing the influence of biased sources.

\section{Evaluation Results}
\label{sec:Result}

\subsection{Survey Prediction Results}

\begin{figure}[t]
    \centering
    \includegraphics[width=0.8\linewidth]{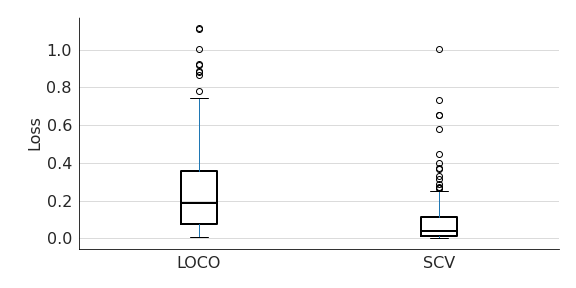}
    \captionsetup{font=small}
    \caption{Boxplots of leave-one-country-out (LOCO) and same-country-validation (SCV) scenarios for predicting survey response favorability towards the United States. Loss is defined as KL divergence of the test sample prediction with the actual survey response data from PEW. The figure demonstrates that using data from the same country in the training set results in significantly lower prediction error.}
    \label{fig:loss_box}
\end{figure}

The results of our evaluation demonstrate the efficacy of our quotation-based modeling approach in predicting survey response favorability towards the United States. Our analysis in Figure~\ref{fig:loss_box} compares the leave-one-country-out (LOCO) and same-country-validation (SCV) scenarios using boxplots to visualize the distribution of test data loss. As per our methodology, loss is measured by the sum of KL divergence of the test sample prediction with the actual survey response data from PEW. The boxplots reveal that the median loss of LOCO is 0.19, while that of SCV is 0.04. The upper percentile of loss for LOCO is 0.35, while for SCV, it is only 0.11. These findings demonstrate that using data from the same country in the training set results in lower prediction error. This is a crucial observation that can aid future studies in optimizing their training data selection to improve model performance.

\begin{figure}[t]
    \centering
    \begin{subfigure}{0.45\textwidth}
    \centering
    \includegraphics[width=\linewidth]{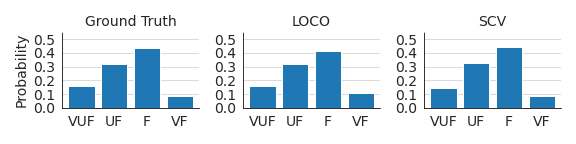}
    \captionsetup{font=small}
    \caption{Australia, 2019, LOCO $\mathcal{L}=0.005$, SCV $\mathcal{L}=0.001$}
    \label{fig:austarila2019}
\end{subfigure}
\begin{subfigure}{0.45\textwidth}
    \centering
    \includegraphics[width=\linewidth]{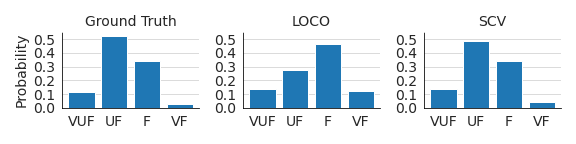}
    \captionsetup{font=small}
    \caption{Germany, 2017, LOCO $\mathcal{L}=0.24$, SCV $\mathcal{L}=0.01$}
    \label{fig:germany2017}
\end{subfigure}
\begin{subfigure}{0.45\textwidth}
    \centering
    \includegraphics[width=\linewidth]{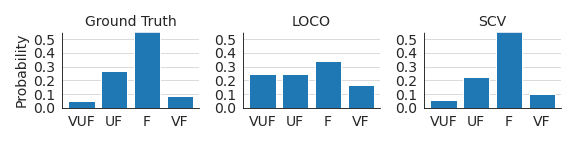}
    \captionsetup{font=small}
    \caption{Hungary, 2018, LOCO $\mathcal{L}=0.32$, SCV $\mathcal{L}=0.01$}
    \label{fig:hungary2018}
\end{subfigure}
\captionsetup{font=small}
\caption{Comparison of predicted quotation-based distribution for LOCO and SCV scenarios in three different countries using KL divergence loss~($\mathcal{L}$). The labels VUF, UF, F, VF correspond the labels in PEW as very unfavorable, somewhat unfavorable, somewhat favorable, and very favorable, respectively. Ground truth data is extracted from PEW dataset. }
\label{fig:loss_samples}
\end{figure}

Figure~\ref{fig:loss_samples} compares the ground truth and predicted quotation-based distribution of LOCO and SCV scenarios for three countries using the PEW dataset. The figure also displays the loss measure, computed as the KL divergence of the ground truth to the predicted quotation-based scenario. The results in Figure~\ref{fig:austarila2019} indicate a low prediction error for both LOCO and SCV, with loss values less than $0.005$. In Figure~\ref{fig:germany2017}, the SCV scenario performs better than LOCO, as seen by the lower loss value of 0.01, compared to LOCO's loss value of 0.24. Also, for this figure, the SCV scenario performs better than LOCO by accurately predicting the most frequent answer~(somewhat unfavorable), while LOCO fails to do so. Finally, for Figure~\ref{fig:hungary2018}, LOCO predicts the most frequent answer, but with high probability on the other options leading to higher loss, whereas SCV predicts the answer with much lower loss value. The figure provides insights into how well the LOCO and SCV scenarios perform and can be used to compare the results with Figure~\ref{fig:loss_box}, which shows the loss function for the entire dataset.

\subsection{Correlation with number of quotes}
\begin{figure}[t]
    \centering
    \begin{subfigure}{0.23\textwidth}
    \centering
    \includegraphics[width=\linewidth]{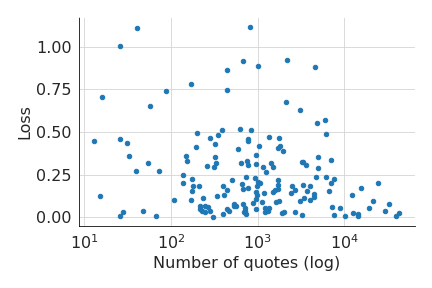}
    \captionsetup{font=small}
    \caption{LOCO}
    \label{fig:corr_loco}
\end{subfigure}
\begin{subfigure}{0.23\textwidth}
    \centering
    \includegraphics[width=\linewidth]{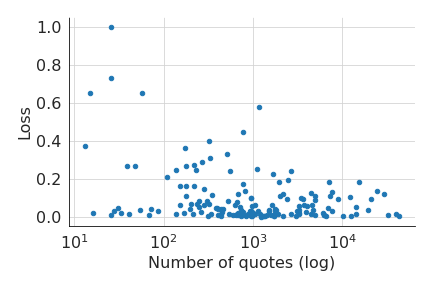}
    \captionsetup{font=small}
    \caption{SCV}
    \label{fig:corr_scv}
\end{subfigure}
\captionsetup{font=small}
\caption{\small{Relationship between loss and the number of quotes for each sample, using LOCO and SCV scenarios, indicating a significant correlation between the number of quotes and the loss.}}
\label{fig:corr}
\end{figure}

This section presents Figure~\ref{fig:corr} depicting the relationship between the loss and the number of quotes for each sample in our test set. Specifically, each sample corresponds to a particular year of a given country. The x-axis of the figure is plotted in a logarithmic scale to represent the number of quotes. The figure includes Figure~\ref{fig:corr_loco} and \ref{fig:corr_scv} representing the results obtained using the LOCO and SCV scenarios, respectively.

To determine the correlation coefficient between the loss and the number of quotes, we used the Pearson Correlation Coefficient and applied a statistical study to the correlation. The null hypothesis that was tested is that there is no significant correlation between the loss and the number of quotes. The p-value obtained from the correlation coefficient for both LOCO and SCV scenarios was less than 0.05, which indicates that the null hypothesis can be rejected. Therefore, we conclude that there is a significant correlation between the number of quotes and the loss, and that having more quotations from a particular country in a specific year can lead to lower loss and improved prediction accuracy for favorability toward the United States.

As shown in Figure~\ref{fig:corr}, the results demonstrate the importance of the number of quotes in predicting survey response, and highlight the advantages of using the SCV over the LOCO scenario. Furthermore, it is observed that the pattern in Figure~\ref{fig:corr} becomes more pronounced after 1000 samples, a phenomenon consistent with fundamental machine learning principles: the scarcity of data tends to increase variance and amplify the overall error.


\subsection{Prediction of Survey Response for Non-PEW Countries}
\begin{figure}[t]
\centering
\includegraphics[width=\linewidth]{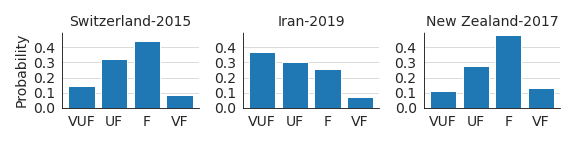}
\captionsetup{font=small}
\caption{Predicted Favorability towards the United States for Three Non-PEW Countries based on Quotation-based Modeling. The labels VUF, UF, F, VF correspond to the labels in PEW as very unfavorable, somewhat unfavorable, somewhat favorable, and very favorable, respectively.}
\label{fig:time_tunnel_independent}
\end{figure}

We evaluate the performance of our quotation-based model on three countries, namely Switzerland, Iran, and New Zealand, which were not included in the PEW research. The model is trained on the extracted quotes from the reference quotebank dataset with specific keywords related to favorability towards the United States. The prediction of the model for these three countries is shown in Figure~\ref{fig:time_tunnel_independent}.

The model predicts that Switzerland and New Zealand have a positive favorability towards the United States, with a label of 'F', while Iran has a negative favorability, with a label of 'VUF'. It should be noted that since we do not have any data from these three countries in the PEW dataset, we need to use LOCO scenario, and we cannot compare our predictions with the actual survey results. This demonstrates the potential of our quotation-based model to be used for predicting survey response in countries that are not covered in existing surveys.


\section{Conclusion}
\label{sec:Conclusion}

In conclusion, our study proposes a novel approach for predicting survey response by analyzing quotations related to the survey question using machine learning techniques. The proposed method involves five main steps, including data collection, sentiment analysis, bias reduction, nationality extraction, and prediction using a KNN regressor. We compared the leave-one-country-out (LOCO) and same-country-validation (SCV) scenarios and found that the SCV scenario outperforms LOCO in terms of median loss. Our results also emphasize the importance of the number of quotes in predicting survey response. Also, this work demonstrates the effectiveness of our quotation-based model for predicting survey response even in countries not included in the PEW research. Our study has the potential to significantly reduce the cost and time required for surveys while providing accurate insights into public opinion towards the United States as a case study.

\section*{Acknowledgment}
We would like to acknowledge Valentin Hartmann and Professor Robert West for valuable feedback, which provided the foundation and knowledge necessary for this research.
\bibliographystyle{IEEEtran} 
\bibliography{example}

\end{document}